\documentclass[]{aa}
\usepackage{amsmath,amssymb,wasysym,stmaryrd,enumerate,longtable,fancyhdr,calc}
\usepackage{floatflt,graphics,calc,epsfig,lscape,natbib,times}
\usepackage[T1]{fontenc}
\usepackage{mathptm}
\newcommand{\tmop}[1]{\operatorname{#1}}

\newcommand{\rad}{rad~m$^{-2}$}
\newcommand{\dg}{$^{\circ}$}

\begin{document}
\title{Magnetic field near the central region of the Galaxy: Rotation measure
of extragalactic sources}
\titlerunning{Magnetic field near the Galactic centre region}
\author{Subhashis Roy\inst{1, 2} \and A. Pramesh Rao\inst{2} \and Ravi Subrahmanyan
\inst{3}}
\institute{ASTRON, P.O. Box 2, 7990 AA Dwingeloo, The Netherlands.
\email{roy@astron.nl}
\and National Centre for Radio Astrophysics (TIFR), Pune University Campus, \\
Post Bag No.3, Ganeshkhind, Pune 411 007, India.
\email{roy@ncra.tifr.res.in} \and
Raman Research Institute, C. V. Raman Avenue, Sadashivanagar, Bangalore 
560 080, India} 
\date{}

\abstract
{}
{We determine the properties of the Faraday screen and the magnetic field near
the central region of the Galaxy.} {We measured 
the
Faraday rotation measure (RM) towards 60 background extragalactic source
components through the $-6^{\circ}<l<6^{\circ}$, $-2^{\circ}<b<2^{\circ}$
region of the Galaxy using the 4.8 and 8.5 GHz bands of the ATCA and VLA. Here
we use the measured RMs to estimate the systematic and the random components of
the magnetic fields.} {The measured RMs are found to be mostly positive for the
sample sources in the region.
This is consistent with either a large scale bisymmetric spiral magnetic
fields in the Galaxy or with fields oriented along the central bar of the
Galaxy. The outer scale of the RM fluctuation is found to be about 40 pc,
which is much larger than the observed RM size scales towards the non thermal
filaments (NTFs). The RM structure function is well-fitted with a power law index
of 0.7$\pm$0.1 at length scales of 0.3 to 100 pc. If Gaussian random processes
in the ISM are valid, the power law index is consistent with a two dimensional
Kolmogorov turbulence.  If there is indeed a strong magnetic field within
$\sim$1\dg\ (radius 150 pc) from the GC, the strength of the random field in the
region is estimated to be $\sim$20 $\mu$G.} 
{ Given the highly turbulent magnetoionic ISM in this region, the strength
of the systematic component of the magnetic fields would most likely be close
to that of the random component. This suggests that the earlier estimated
milliGauss magnetic field near the NTFs is localised and does not pervade the
central 300 pc of the Galaxy.}

\maketitle

\section{Introduction:}
\label{rm.intro}
Magnetic fields are widely recognised as playing an important role in the
evolution of supernova remnants, in star formation, overall structure of ISM,
cosmic ray confinement and non-thermal radio emission. This is especially true
in central region of the Galaxy, where magnetic fields could be strong enough
to be significant in the dynamics and evolution of the region \citep{BECK1996}.
A relatively high systematic magnetic field in the Galactic centre (GC)
region was believed to be responsible for the creation and maintenance of the
unique non thermal filaments (NTFs) (Morris et al. 1996, and references
therein).  Therefore, it is important to measure the magnetic-field geometry
and strength near the central part of the Galaxy.

Other than the central 200 pc of the GC, no systematic study has been made in
the past to measure the magnetic fields in the inner 5 kpc region of the Galaxy
\citep{DAVIDSON1996}. Recently, \citet{BROWN2007} have surveyed the 4th
quadrant of the Galaxy up to $l$=358\dg\ through Faraday RMs, but their
observations do not target the central kpc of the Galaxy.  The earlier
estimates of magnetic fields within the central 200 pc of the Galaxy were based
mainly on observations of the non-thermal filaments, and the measured Faraday
rotation measure (RM) towards these NTFs were found to be $\sim$1000 \rad\
\citep{YUSEF-ZADEH1987b, ANANTHARAMAIAH1991,
GRAY1995,YUSEF-ZADEH1997,LANG1999b}.  Since magnetic pressure in these NTFs
appears to overcome turbulent ISM pressure (otherwise, the NTFs would have bent
due to interaction with molecular clouds), \citet{YUSEF-ZADEH1987a} derived a
magnetic-field strength of about 1 milliGauss within these NTFs.
Moreover, the high magnetic field in the region is required to be ubiquitous. 
Otherwise, in regions where there is no molecular cloud around NTFs,
magnetic pressure within these structures would be much higher than outside.
This will cause the NTFs to expand at Alfven speed and decay at a time scale
($\sim$300 years) that is likely to be much less than their formation time
scales \citep{MORRIS1998}.  This implies that if the NTFs are static
structures, the magnetic field in the region must be ubiquitous (Morris et al.
1996).

Earlier measurements of direction of magnetic fields within the NTFs have
shown it to be oriented along their length. Since all the well known NTFs found
within a degree of the GC are oriented almost perpendicular to the Galactic
plane, it suggests the field lines in the surrounding ISM are also
perpendicular to the Galactic plane (Morris et al. 1996 and the references
therein). In addition, the NTF Pelican (G358.85+0.47) \citep{LANG1999a} located
about a degree from the GC is found to be almost parallel to the Galactic
plane. This indicates that the field lines change their orientation from being
perpendicular to parallel to the plane beyond a degree from the GC, which is
typically observed in the rest of the Galaxy.  However, we note that if the NTFs are
manifestations of peculiar local environments \citep{SHORE1999}, inferences
drawn from these observations can be misleading.  With the recent discovery of
many new fainter filamentary structures in the GC region oriented quite randomly
with the Galactic plane \citep{NORD2004}, serious doubts are cast on
the orientation of the magnetic field and its ubiquitous nature near the GC.

Zeeman splitting of spectral lines can directly yield the magnetic field in a
region.  However, this method is known to be sensitive to small-scale fields,
and therefore high magnetic field strengths in a small region anywhere along
the line-of-sight (LOS) can indicate a high magnetic field, which is not
representative of the average. Therefore, past estimates of milliGauss magnetic
fields based on Zeeman splitting \citep{SCHWARZ1990, KILLEEN1992,
YUSEF-ZADEH1996, YUSEF-ZADEH1999} of HI or OH lines towards the GC could have
resulted from local enhancement of field (\textit{e.g.}, near the cores of
high-density molecular clouds). To measure any systematic magnetic field in the
region, it is necessary to use an observational technique that is sensitive to
large-scale fields. To avoid manifestations of favourable local environments,
Galactic objects should not be used for this purpose.

Faraday rotation measure is the integrated LOS magnetic field
weighted by the electron density 
\begin {equation}
\label{rm.eqn.1}
RM=0.81 \times\int n_e B_{\|}dl, 
\end{equation}
where, RM is rotation measure expressed in rad m$^{-2}$, n$_e$ is the
electron density expressed in cm$^{-3}$, B$_{\|}$ is the LOS
component of the magnetic field in $\mu$G, and the integration is carried out
along the LOS, with distance expressed in parsec. If a model for the
electron density is available, observations of RM towards the extragalactic
sources seen through the Galaxy can be used to estimate the average magnetic
fields in the ISM. A large number of studies of the Galactic magnetic fields
have already been made using RM towards the extragalactic sources
\citet{SIMARD-NORMANDIN1980,FRICK2001,CLEGG1992}. Similar studies have also
been carried out towards pulsars \citep{RAND1989, RAND1994, HAN1994}.  These
studies have shown that there is one field reversal within and one
outside the solar circle, while two more reversals have been suggested by
\citet{HAN1999}. These reversals could be explained by invoking either the
bisymmetric spiral model \citep{SIMARD-NORMANDIN1980, HAN1999} or 
a ring model, where the direction of the field lines reverses in each ring
\citep{RAND1989}. However, there has been no systematic observation of RM
towards extragalactic sources seen through the GC region.

We systematically studied RM properties of 60 extragalactic sources seen
through the central $-$6$^{\circ}<l<$6$^{\circ}$, $-$2$^{\circ}<b<$2$^{\circ}$
region of the Galaxy. 
The angular scale over which the magnetoionic medium is coherent near
the NTFs has been estimated as $\sim $10$^{''}$ (Gray et al. 1995,
Yusef-Zadeh et al.  1997).  Therefore, to avoid any beam depolarisation
introduced by the ISM of the Galaxy, our observations were made with the
higher resolution configurations of these telescopes, so that the synthesised
beam sizes are considerably smaller than the coherence scale length of the
Faraday screen near the GC.  Preliminary results of these observations were
published earlier in \citet{ROY2003ANS} and \citet{ROY2004BASI}.  In 
\citet{ROY2005} (henceforth Paper I), we described the sample
sources, the observations and data analysis and then determined their spectral
indices, polarisation fraction, RM, and the direction of their intrinsic
magnetic field.  In this paper, we interpret the RM observations.  In
Sect.~\ref{rm.results}, we provide a graphic representation of the
measured RMs (see Paper-I), while interpretation of the results is
described in Sect.~\ref{rm.discussion}. The conclusions are presented in
Sect.~\ref{rm.summary}.

\section{Results}
\label{rm.results}

\subsection{Features in the Faraday screen near the GC}

\begin{figure*}
\centering
\includegraphics[width=0.9\textwidth,clip=true,angle=0]{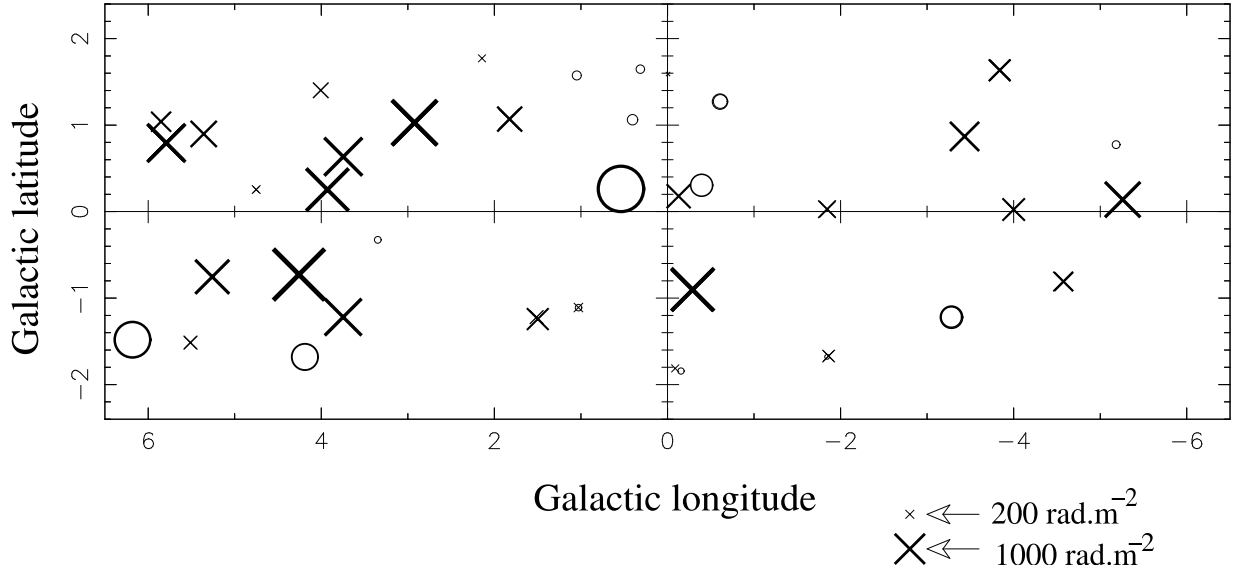}
\caption{Measured Faraday RMs towards the polarised sources in Galactic
co-ordinate. The positive values are indicated by `cross (X)' and negative
values by `circle (O)'. Size of the symbols increase linearly with $|$RM$|$.}
\label{gc.rm.plot}
\end{figure*}

In Fig.~\ref{gc.rm.plot}, we plot RMs of 60 polarised components (including
2 secondary calibrators), which conform to the criteria given in Paper I (i.e.,
reduced $\chi^2$ of the polarisation angle vs. frequency fit less than or equal
to 4.6, depolarisation fraction between 4.8 and 8.5 GHz higher than or equal to
0.6 and the source is outside the Galaxy). This figure shows our measured RMs
divided into four quadrants according to the signs of Galactic longitude and
latitude. In the rest of this paper, we define quadrant A when $l$ and $b$ are
both positive, quadrant B when $l$ is negative but $b$ is positive, quadrant C
when both $l$ and $b$ are negative and quadrant D when $l$ is positive but $b$
is negative. 
The region is dominated by positive RMs, as observed towards most of the
sources in both positive and negative Galactic longitude.  The observed RMs
towards sources with $|b| \le$ 1.5\dg\ are quite high $\sim $1000 \rad. Such
high RMs have been measured towards extragalactic sources at low Galactic
latitudes (45\dg $< l <$93\dg, and $|b| <$ 5\dg) by \citet{CLEGG1992} and are
due to passage of radio wave through large path lengths along interstellar
medium.
These results are consistent with positive RMs observed near $l$=$-$5\dg\ by
\citet{BROWN2007}, which lies near the edge of their survey.

In general, magnetoionic media responsible for the RMs have structures at
different length scales. Reversal of sign of the RM over angular scales of a
few degrees shows the existence of a random component of the magnetic field.
We explore this through the structure function analysis of the RMs and then
identify systematic features in the data.

\subsubsection{The structure function analysis of RMs}

\begin{figure}
\centering
\includegraphics[width=0.5\textwidth,clip=true,angle=0]{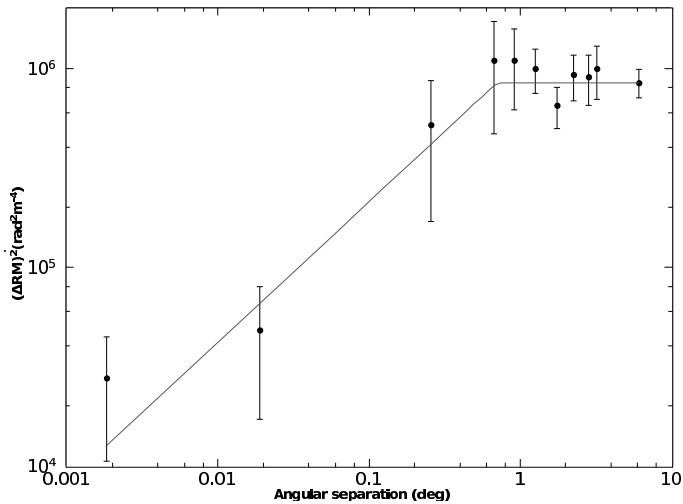}
\caption{The structure function of the measured RMs. See text for the fit.}
\label{structure-function}
\end{figure}

Variations in RM over an angular scale of $\Delta \theta$ can be described by
the RM structure function $D(\Delta \theta) = <[ RM(\theta) - RM(\theta +
\Delta \theta)]^2 >$. The structure function is measured by computing the
expectation value of the squared differences of the RM among all pairs of
sources within a particular range of angular separation ($\Delta \theta$). We
binned the data with $\Delta \theta$ from 
0.0\dg\ to 0.005\dg, 0.005\dg\ to 0.1\dg, 0.1\dg\ to 0.33\dg, 0.33\dg\ to
0.6\dg, and then up to 1.0\dg\ in bin widths of 0.2\dg. From 1.0\dg\ to 4.0\dg,
we binned angular separations with bin widths of 0.5\dg, and from 4.0\dg to
10\dg\, a single bin was used. The RM structure function in each bin is plotted
at the location of the median angular separation of sources in that bin in
Fig.~\ref{structure-function}. The errorbars in the plot were estimated by a
statistical method called `Bootstrap' \citep{EFRON1977}.
Figure~\ref{structure-function} shows that
the structure function appears to saturate at beyond $\sim$0.7\dg; therefore,
we fitted a power law $f(\Delta \theta)=A \times \Delta \theta ^n$ for
$\Delta \theta \le$ 0.7\dg, and then $f(\Delta \theta)$ is held fixed at its
value for $\Delta \theta$=0.7\dg. The function [$f(\Delta \theta)$] is
well-fitted to the data (reduced $\chi ^2$ 0.47), and is shown in
Fig.~\ref{structure-function}. From the fit, A is found to be
1.1$\pm$0.1$\times$ 10$^6$, and the power law index ($n$) is 0.7$\pm$0.1.

The outer scale of the structure function is defined as the length scale at
which the structure function attains half of its maximum value
\citep{RICKETT1988}. From the data, we estimate the outer scale to be about
0.3$^{\circ}$ or 40 pc for the screen if located at the distance of GC, 8.0 kpc
away.  This indicates the RMs of sources lying within an angular distance of
$<$0.3$^{\circ}$ are likely to be correlated.  Therefore, while determining
statistical quantities in this paper, we ensure independent measurements by
considering the RMs of source components located at least beyond 0.2$^{\circ}$
from each other. There are a total of 38 source components that conforms to
this criterion.

Given the extreme conditions in the ISM close to the GC, statistical
properties of the medium in this region could be different from that of its
immediate surroundings. To check for any change in outer scale of RM for
sources seen within 1.1\dg\ ($\sim$~150 pc) of the GC, we carried out the above
analysis for 6 sources seen through the region.  The structure function of
these sources with angular separations less than 1.1\dg\ is 9.5$\pm$5.5
$\times$10$^5$ \rad, and is 3.8$\pm$1.6 $\times$10$^6$ \rad\ for sources with
angular separations between 1.1\dg\ to 2.1\dg. This shows the RMs changed by
more than 1.7 times the effective error, indicating that the structure function
of RMs is not saturated for angular separations of less than a degree
($>$90\% confidence). While the significance of this result is not very high
due to the small number of sources in the sample, we adopt the simplest model
and assume the outer scale in the inner 1.1\dg\ region is comparable to 40 pc
determined from the full sample.

\subsubsection{Large-scale pattern in the RMs data}
\begin{figure}
\begin{minipage}{0.45\textwidth}
\centering
\includegraphics[width=0.75\textwidth,clip=true,angle=270]{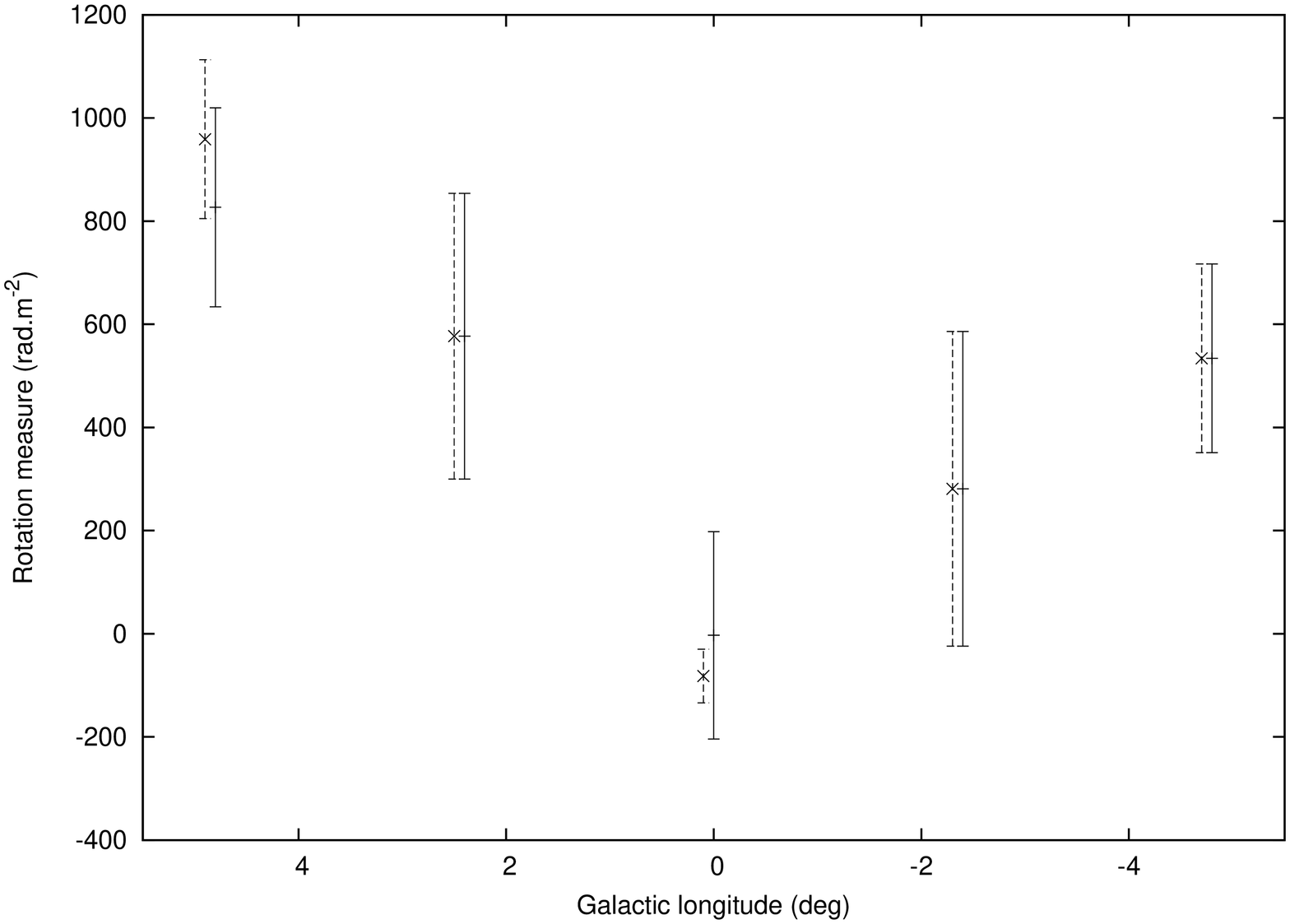}
\caption{Plot of RMs as a function of Galactic longitude. The values shown are
averaged over 2.4 degree bins in longitude. Data points with solid error bars
are before flagging, and data after flagging are displayed by dashed errorbars
with axis shifted by $-$0.1\dg\ to what is displayed along the axis at the
bottom. Details of flagging are described in the text.}
\label{l.rm.grid}
\vspace{0.5cm}
\end{minipage}
\hfill
\begin{minipage}{0.45\textwidth}
\centering
\includegraphics[width=0.75\textwidth,clip=true,angle=270]{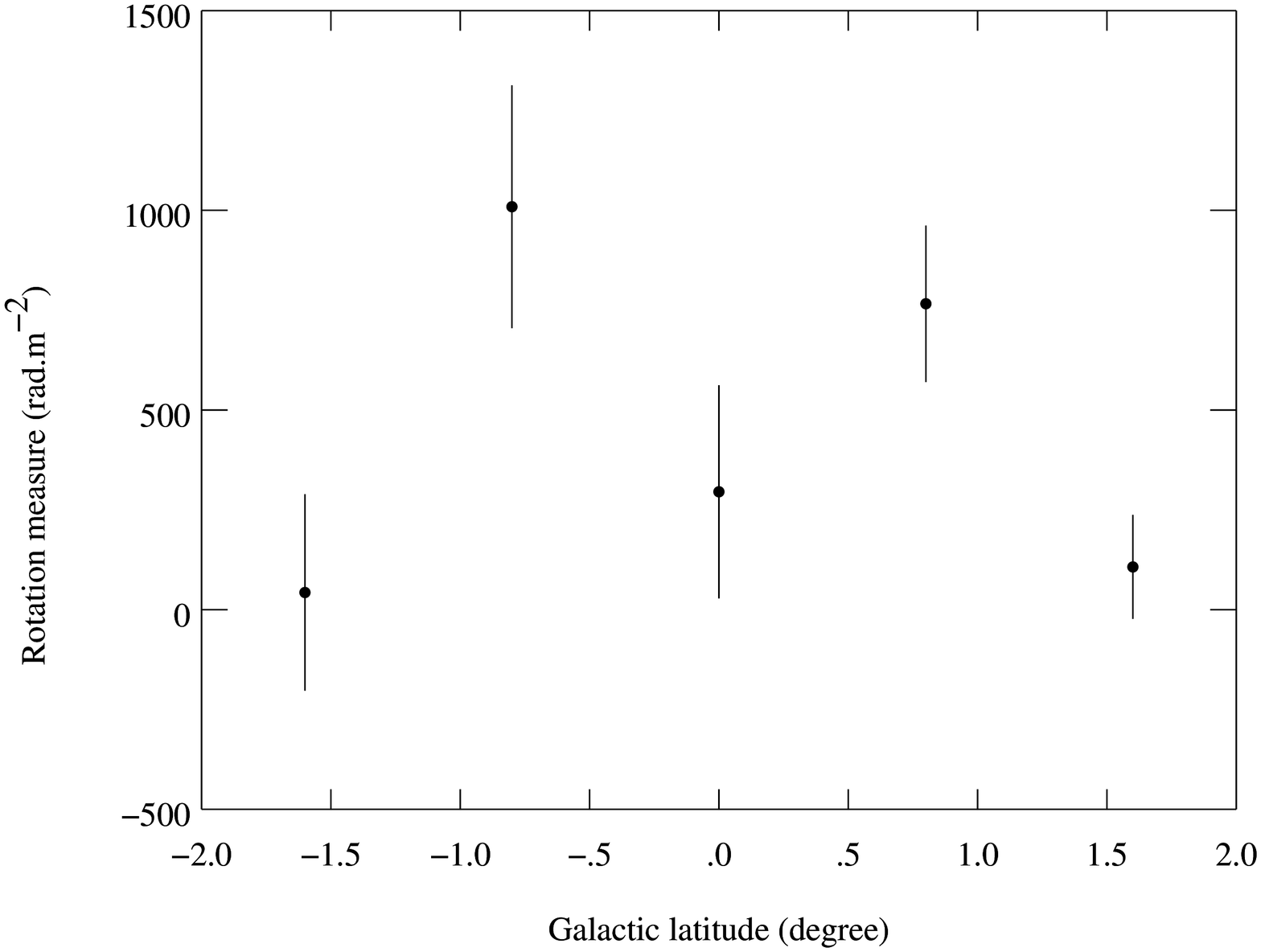}
\caption{Plot of RMs as a function of Galactic latitude. The values shown are
averaged over 0.8 degree bins.}
\label{b.rm.grid}
\end{minipage}
\end{figure}

In Fig.~\ref{gc.rm.plot}, we notice the dominance of sources with positive RMs.
Following the criteria given above for sources with uncorrelated RMs, we find
the mean RM from the data to be 413 $\pm$115 \rad, and the median is 476 \rad.
The mean RM in quadrant A of Fig.~\ref{gc.rm.plot} is 488$\pm$204, 396$\pm$294
in quadrant B, 354$\pm$170 in C and 350$\pm$323 in D.  To study the systematic
behaviours of this large-scale field, we divided the observed region in 
several bins along the Galactic longitude and latitude such that a reasonably
large number of sources remain in each bin to yield meaningful
statistical properties (mean, rms) of RMs in these bins. Therefore, we 
selected 5 bins along the Galactic longitude, each 2.4\dg\ wide resulting in
$\sim$8 sources per bin, and the resulting statistical properties of RMs
from each of these bins is tabulated in Table.~1.  Similarly, in Table.~2 we
have tabulated statistical properties of RMs along Galactic latitude divided
in 5 bins of width 0.8\dg.  Average RMs along Galactic longitude and
latitude are plotted in Fig.~\ref{l.rm.grid} (shown with solid error-bars) and
Fig.~\ref{b.rm.grid}, respectively.  Figure~\ref{l.rm.grid} shows that the
average RM of sources located within $|l| <1$\dg\ is significantly less than
that of sources located beyond $|l| >3$\dg.  

To remove possible small-scale variations in RM due to LOS HII regions or
supernova remnants \citep{MITRA2003}, thereby getting a clearer picture of the
large-scale field in the region, we used the following method.  We estimate the
mean and rms RM of sources located in each bin in Table.~1. If measured
RM of any source within a bin deviates from the mean in that bin beyond 1.7
times the rms ($\le$10\% probability for Gaussian distributed errors), that RM
is rejected (flagged) and the mean and rms RM in that bin is recomputed.  This
process is repeated till there is no source outside the flag limit. Since
there are only a few sources per bin, flagging the highly deviant data
points in a bin would reduce the measured rms as compared to the real rms in
the data.  However, in a majority of cases, it results in a drop in measured
rms of only $\sim$25\%, and the probability of a decrease in rms noise by a
factor of 4 is $<$10\%.
Five out of 38 sources were rejected as a result of the flagging by this
method, and the minimum number of sources in any of the bins after flagging
were 4.  The resulting mean and rms RM values in each bin are tabulated in
Table.~1.  We note that most of the flagging was in the central bin with
$-$1.2\dg $<l$ $<$ 1.2\dg, where 4 out of 11 sources were flagged, 
and the rms RM of sources decreased by almost a factor of 5 after flagging
($\sim$5\% probability with Gaussian random noise), indicating a
significant small-scale structure (non-Gaussian errors) in the Faraday screen
towards this region. The resulting distribution of average RM is plotted in
Fig.~\ref{l.rm.grid} using dashed errorbars, and to make the symbols visible,
the X-axis of this plot is shifted by $-$0.1\dg\ from what is shown at the
bottom. This 
shows that the average RM tends to zero near $l$=0\dg. We applied the same
procedure for RMs of sources located in each of the bins in Table.~2, and the
resulting mean and rms RM values after flagging in each bin are tabulated
there. No significant change in mean or rms RM is noticed after flagging in
this case.

\begin{table*}
	\caption{Rotation measure of sources binned along Galactic 
	longitude}
  \begin{tabular}{|c|c|c|c|c|c|c|c|}
\hline
    Bin & Range & No. of & Mean & Rms on & No. of & Mean RM
    & Rms on mean \\
    No. & in $l$ & sources & RM & mean RM & sources &
    after sources &  RM after\\
    &  &  &  &   & flagged & flagged & flagging\\
    & (deg) &  & (rad.m$^{- 2}$) & (rad.m$^{- 2}$) & 
    & (rad.m$^{- 2}$) & (rad.m$^{- 2}$)\\
\hline
    1 & $-$6.0 to $-$3.6 & 5 & 534 & 183 & 0 & 534 & 183\\
    2 & $-$3.6 to $-$1.2 & 4 & 281 & 305 & 0 & 281 & 305\\
    3 & $-$1.2 \ \ to \ 1.2 & 11 & $- 3$ & 201 & 4 & $-$82 & 52\\
    4 & 1.2 \ \ to \ 3.6 & 5 & 577 & 277 & 0 & 577 & 277\\
    5 & 3.6 \ \ to \ 6.0 & 12 & 827 & 193 & 1 & 959 & 154 \\

\hline
  \end{tabular}
\end{table*}

\begin{table*}
	\caption{Rotation measure of sources binned along Galactic 
	latitude}
  \begin{tabular}{|c|c|c|c|c|c|c|c|c|}
\hline
    Bin & Range & No. of & Mean & Rms on & No. of & Mean RM & Rms on mean \\
    No. & in $l$ & sources & RM & mean RM & sources & after sources &  RM after\\
    &  &  &  &   & flagged & flagged & flagging\\
    & (deg) &  & (rad.m$^{- 2}$) & (rad.m$^{- 2}$) & 
    & (rad.m$^{- 2}$) & (rad.m$^{- 2}$)\\
\hline
    1 & $-$2.0 to $-$1.2 & 8 & 43 & 246 & 1 & -129 & 203 \\
    2 & $-$1.2 to $-$0.4 & 5 & 1009 & 304 & 0 & 1009 & 304 \\
    3 & $-$0.4 \ \ to \ 0.4 & 9 & 295 & 267 & 1 & 478 & 220 \\
    4 & 0.4 \ \ to \ 1.2 & 9 & 766 & 196 & 0 & 766 & 196 \\
    5 & 1.2 \ \ to \ 2.0 & 7 & 107 & 130 & 0 & 107 & 130 \\
\hline
  \end{tabular}
\end{table*}

\section{Discussion}
In the previous section we identified a largely positive RM towards background
sources, the correlation length of which is about 40 pc. The RMs averaged in
bins along the Galactic longitude was found to decrease near $l$=0\dg.  In this
section, we identify the location and properties of the Faraday screen that
is responsible for the above.  Then, using a plausible model of the electron
density distribution near the GC, we investigate the nature of the magnetic
fields (comprised of systematic and random components) in the region.

\label{rm.discussion} 
\subsection{Location of the Faraday screen}
\subsubsection{Small-scale structures in the Faraday screen and intrinsic RMs
of sources} 

Differences in RMs seen along different LOS could occur from either (i) a
geometrical effect or (ii) change in the property of the Faraday screen.  The
structure function due to a perfectly uniform Faraday screen will have a
measurable geometrical component simply because of the change in the LOS
component of the field with change in the $l$ and $b$. An observer embedded in an
extended homogeneous medium with uniform magnetic field approaching from an
arbitrary angle $\theta_0$ sees a rotation measure RM$_0 cos(\theta-\theta_0)$
\citep{CLEGG1992}, where RM$_0$ is the RM towards $\theta_0$. 
However, over the observed longitude range, the contribution from the variation
in the `cosine' term is much less than what is observed in
Fig.~\ref{structure-function}.  Therefore, we do not consider the geometrical
effect any further.

The measured RMs towards the sources could have a significant intrinsic
contribution from a magnetoionic medium local to the sources. However, in this
case, intrinsic RMs towards different sources will be uncorrelated.
Consequently, differences in RMs for unrelated sources will persist regardless
of their location on the sky plane, and will not approach zero when their
angular separation tends to zero.  However, in Fig.~\ref{structure-function}, 
we find the RM structure function tends to zero at zero angular separation and
increases smoothly with source angular separations. This shows intrinsic RMs
can be neglected, and the measured RMs have an interstellar origin (plasma
turbulence) within our Galaxy.

\subsubsection{RM contribution from the Galactic disk}
To explain the observed magnetic field orientation in our Galactic disk, two
models of magnetic fields, the ring \citep{RAND1989} and the bisymmetric spiral
\citep{SIMARD-NORMANDIN1980} are widely used. However, both of these models
predict that the LOS RM contribution from the Galactic disk is quite small when
$|l| <<1$ rad. Ionised gas located close to us along the LOS could, however,
produce a bias on a large angular scale to the observed RMs. In this case,
nearby pulsars seen towards GC will also show such correlated RMs. We 
searched for pulsar RMs located in our survey region from ATNF Pulsar Catalogue
\citep{Manchester2005} and found 7 pulsars with measured RMs that are located
closer than the GC. Their distance as estimated from their dispersion measure
\citep{TAYLOR1993} varies from 1.5 kpc to 7.7 kpc with a median value of 3.5
kpc.  Mean RM of these sources is $-$7$\pm$46 \rad.  Since the mean RM is quite
small, any Faraday screen affecting our sample has to be located at least
beyond the median distance of these pulsars. Moreover, the linear size of an
object at this median distance of 3.5 kpc with angular size of our survey will
be $\sim$300 pc. Objects known to produce significant RMs (e.g., HII regions,
supernova remnants) are typically much smaller than the above size scale.
Therefore, no single nearby object has significantly biased the RMs,
so we believe the central few kpc region of the Galaxy is responsible for the
observed RMs.

\subsection{Magnetic field near the GC}
Faraday rotation being the LOS integral of the product of the
magnetic field with the electron density, changes in electron density or the
magnetic field strengths or a change in the direction of the magnetic field
vector can contribute variations in the observed RM. To separate the
contribution of these effects, we first discuss the available models of
the electron density distribution and then discuss the large-scale magnetic
field near the central region of the Galaxy.

\subsubsection{Electron density distribution near the GC and strength of
the large-scale magnetic field}
\label{elec.den}
The electron density of the ISM is believed to increase towards the central
region of the Galaxy. Different electron density models are invoked for the
inner Galaxy, central kpc, and the central 100 pc of the Galaxy, which are
discussed below.

\citet{TAYLOR1993} modelled electron density distribution in the Galaxy and
included an inner Galactic component that is considered a ring at a distance
of $\sim$4 kpc from the GC. However, their model does not include a GC
component. 
Over the central few degrees of the GC, \citet{BOWER2001} carried out VLBA
observations of 3 extragalactic sources and report a region of enhanced
scattering covering $\gtrsim$ 5\dg\ in longitude and $\le$ 5\dg\ in latitude.
The measured scattering diameters correspond to about $\sim$300 milli-arcsec at
1 GHz, which is 1.5--6 times the prediction from the \citet{TAYLOR1993} model.  
Using scatter broadening of OH masers in the vicinity of OH/IR stars,
\citet{LANGEVELDE1992} showed that there is a region of high scattering within
30$'$ of the GC.  From free-free absorption measurements, they
suggest the scattering region is at a distance of more than 850 pc from
the GC. Using a likelihood analysis, \citet{LAZIO1998} claim a `hyperstrong'
scattering screen ($n_e \sim$10~cm$^{-3}$) of the same angular
extent (30$'$) towards the GC, but estimated the distance to this screen to be
$133_{-80}^{+200}$~pc from the GC. This model predicts a scattering diameter for
extragalactic sources to be an order of magnitude higher than what is 
observed by \citet{BOWER2001}. However, the extragalactic source G359.87+0.18
\citep{LAZIO1999} is seen through the `hyperstrong scattering' region, but its
scattering size is an order of magnitude lower than predicted from the
`hyperstrong scattering' model. This indicates the screen is patchy
\citet{LAZIO1999}.  An improved version of \citet{TAYLOR1993} model has been
published by \citet{CORDES2002}, where contribution from a GC component
corresponding to the contribution from the central 30$'$ region of the Galaxy
\citep{LAZIO1998} has been added.  However, it does not include any
contribution from the enhanced scattering region observed by \citet{BOWER2001}. 
In our observations, all the objects barring one (G359.87+0.18)  are seen
through the region of enhanced scattering observed by \citet{BOWER2001}.
Therefore, we used their observations to estimate electron density, which will
be used in the rest of the paper. 
If we assume the turbulence scale length of this screen to be the same as that
of the inner Galaxy component of \citet{TAYLOR1993}, \citet{BOWER2001}
scattering measure imply an electron density of about 0.4 cm$^{-3}$. This is
twice of what is estimated from the \citet{TAYLOR1993} model for the inner Galaxy.
The corresponding dispersion measure from the inner 2 kpc of the Galaxy is 800
pc cm$^{-3}$.  From the \citet{CORDES2002} model, we also estimate the dispersion
measure from the rest of the Galaxy along the LOS passing about a degree away
from the GC, which is found to be 800 pc~cm$^{-3}$.  Therefore, the total
dispersion measure towards the inner kpc of the Galaxy is $\sim$1600
pc~cm$^{-3}$, and half of the total dispersion measure originates from the
inner Galaxy component.  It should be noted that at present the dispersion
measure of the inner Galaxy component is uncertain by factor of a few.
Using the above-mentioned dispersion measure of 800 pc~cm$^{-3}$ for the
central 2 kpc of the Galaxy and mean RM of 413 \rad\ (Sect.~\ref{rm.results})
in Eq.~\ref{rm.eqn.1}, the mean LOS magnetic field is estimated to be
0.6~$\mu$G. As this is an LOS average, it should be treated as a lower limit.

\subsubsection{Geometry of the large scale azimuthal magnetic field}
In the presence of various turbulent processes in the GC, any unravelling of
the large-scale field orientation needs to be performed statistically, and here
we consider possible models to explain the results (Sect.~\ref{rm.results}).  

(i) Magnetohydrodynamic model:\\
\citet{UCHIDA1985} proposed this model to explain the Galactic
Centre Lobes (GCL), which are a pair of limb-brightened radio structures of
several hundred parsecs extending from Galactic plane towards positive
Galactic latitudes \citep{SOFUE1984} and seen within the central 1\dg\ of the
Galaxy.  They carried out non-steady axisymmetric magnetohydrodynamic
simulations in which the magnetic field is assumed to be axial at high Galactic
latitudes.  However, due to the differential rotation of dense gas near the
Galactic plane, the field acquires a component along this plane.  This model
predicts an LOS field in quadrants A and C towards the observer (positive RM),
and away from the observer in quadrants B and D. \citet{NOVAK2003} find the
signs of the measured RMs towards the known NTFs to be consistent with the
above prediction.  From our observations, the estimated mean RM towards sources
seen through quadrants A and C is 432$\pm$133 \rad\ and 379$\pm$217 \rad\
through quadrants B and D . Positive RMs in all the quadrants are inconsistent
with their prediction.  However, our sources are observed over a significantly
bigger region around the GC than the NTFs are seen, and the results do not
match the prediction of this model. 

(ii) Ring model: \\
According to this model, magnetic field lines in a galaxy are oriented along
circular rings in the galactic plane.  As discussed in \citet{RAND1989}, such a
geometry arises in galactic dynamo models of the field, in which a symmetric
azimuthal mode is dominant (e.g., \citealt{KRAUSE1987}). Theories involving a
primordial origin of magnetic field also claim to be able to produce ring
fields, but only in the inner regions of galaxies \citep{SOFUE1986}.  In this
model, the LOS magnetic field reverses with the sign of galactic longitude at a
particular galactocentric radius (r). Reversals of the magnetic field as a function
of galactocentric radius are also predicted by this model.  Since both these
predictions are inconsistent with the data (Fig.~\ref{l.rm.grid}), the
ring model is not applicable in this region.

\begin{figure}
\begin{minipage}{0.45\textwidth}
\centering
\includegraphics[width=\textwidth,clip=true]{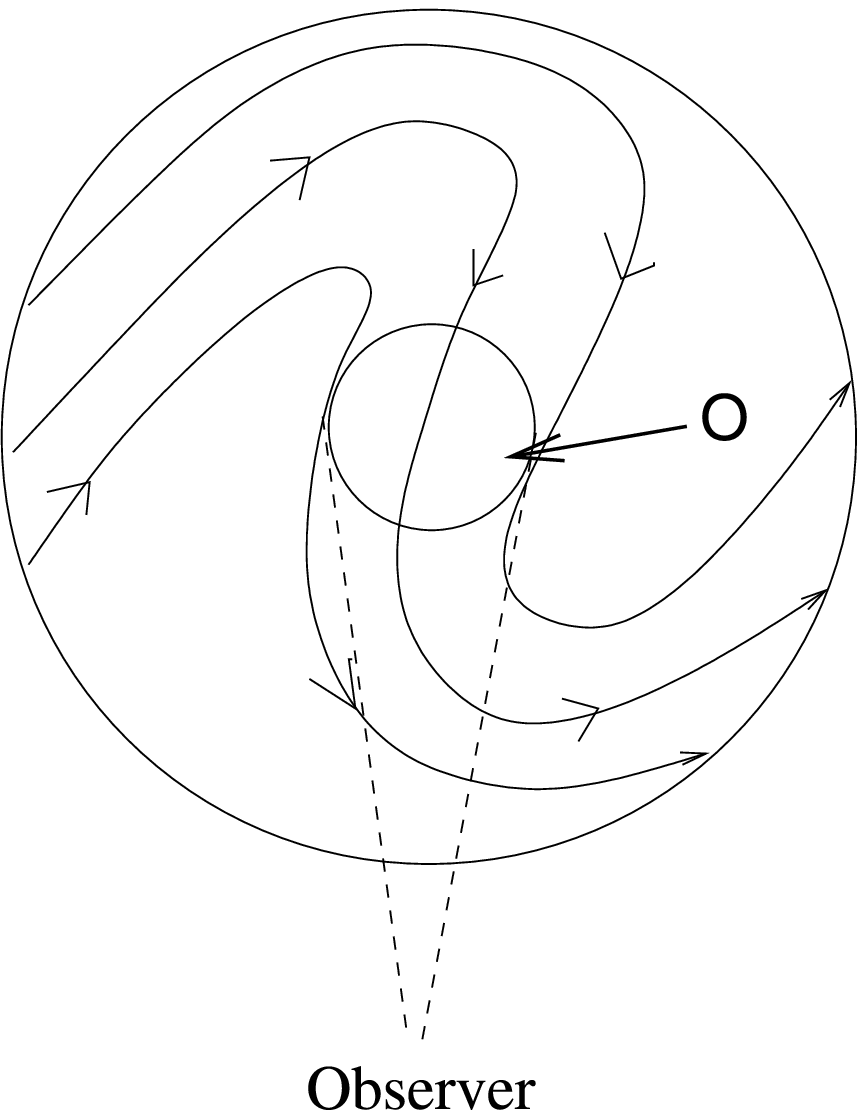}
\caption{A schematic diagram of the bisymmetric spiral structure of magnetic
fields.} 
\label{bisymmetric.spiral}
\end{minipage}
\hfill
\begin{minipage}{0.45\textwidth}
\centering
\includegraphics[width=\textwidth,clip=true,angle=0]{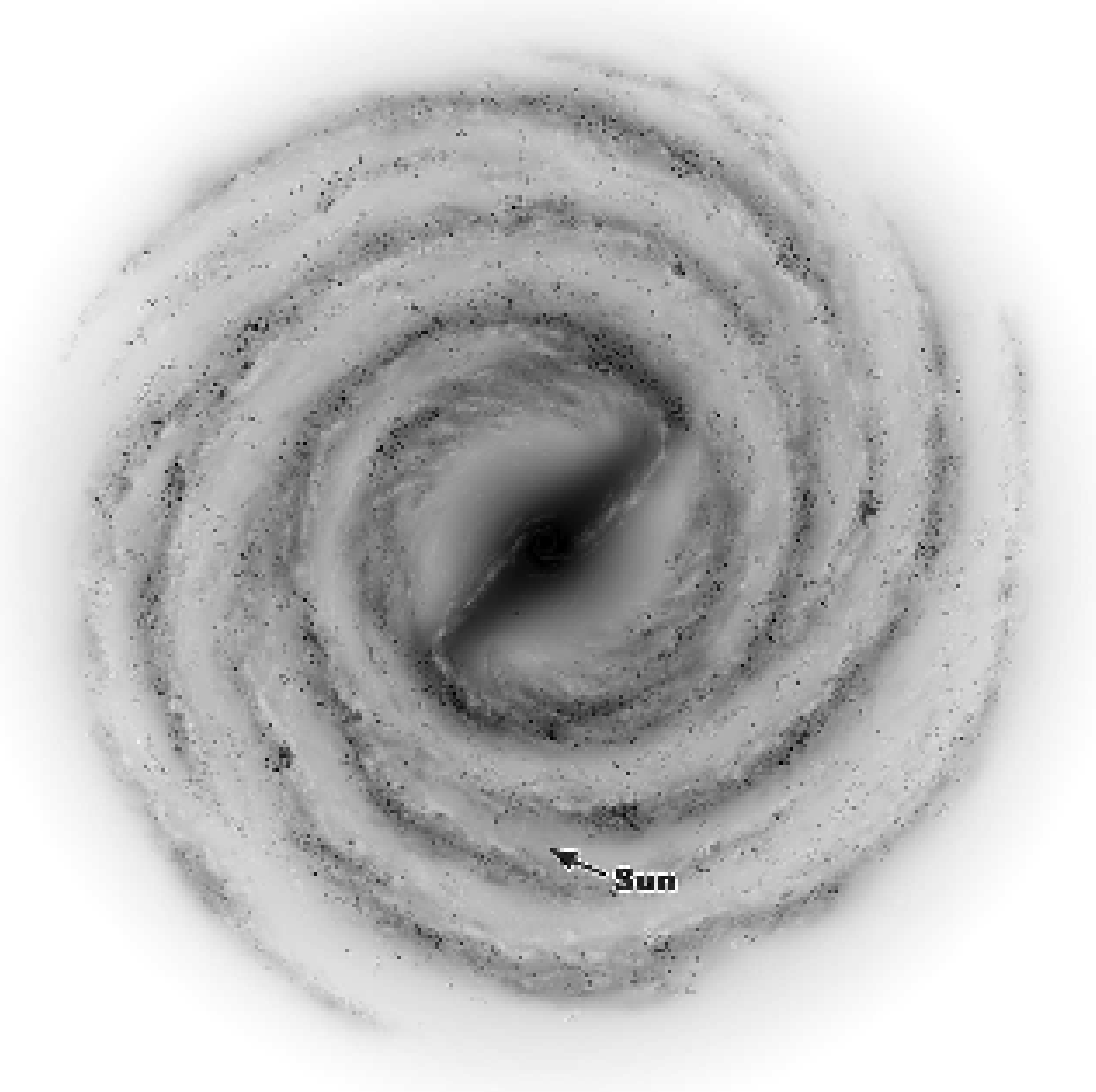}
\caption{A schematic view of the Galaxy from the top of the Galactic plane.
Notice the kpc scale bar within the central 4 kpc from the GC.
}
\label{galaxy.bar}
\end{minipage}
\end{figure}

(iii) Bisymmetric spiral model: \\
\citet{SIMARD-NORMANDIN1980} proposed this model (see also \citet{HAN1999}) to
account for the reversals of magnetic fields with galactocentric distances in
the Galaxy. A schematic diagram of this model is shown in
Fig.~\ref{bisymmetric.spiral}. 
It predicts a positive RMs towards $l$=0\dg, which is what is observed.
Therefore, the prediction from this model near the GC is consistent with our
observations.

(iv) Another plausible configuration of the magnetic field: \\
Magnetic field lines are typically observed to be aligned with large-scale
structures in the Galaxy and beyond. In the central few kpc region of the Galaxy
a bar-like distribution of matter has been suspected for a long time, and
recent Spitzer observations suggest it is oriented at an angle of 44\dg\ with
respect to our LOS \citep{CHURCHWELL2005}. An impression of this from the top
of our Galaxy is shown in
Fig.~\ref{galaxy.bar}\footnote{http://www.spitzer.caltech.edu/Media/mediaimages/sig/sig05-010.shtml}.
A bar in gas distribution in the central region of the Galaxy has also been
claimed \citep{SAWADA2004}. If the magnetic field lines are oriented along this
bar and have a component towards us, then this could explain the positive RMs
observed in all the four quadrants.
We note a decrease in averaged RM near $l=0$\dg\
(Fig.~\ref{l.rm.grid}).  Magnetic fields in the GC region are very likely
anchored to the dense molecular clouds, and within $\sim$1\dg\ of the GC, they
have large random motions, which reduces the magnetic field averaged over the
2.4\dg\ bin centred on the GC.

\subsubsection{Random component of magnetic field}
\label{small.scale.rm}
In this section, we discuss the power spectrum of the magnetoionic ISM and
then estimate the strength of the magnetic field responsible for it.
Small-scale variations in a magnetoionic medium are likely to be related to
electron density fluctuations in ISM, which have been studied through
scattering and scintillation observations \citep{RICKETT1990}. The power
spectrum of electron density irregularities is expressed by 

\begin{equation}
\label{eqn.structure.function}
P(q) = C_n^2 q^{-\alpha}, q_0 < q <q_i
\end{equation}
Where, $q$ is spatial wavenumber, and $\alpha$ is spectral index
\citep{RICKETT1977}. $C_n^2$ is normalisation constant of the electron density
power spectrum. The quantities $q_0$ and $q_i$ represent wavenumbers
corresponding to `outer scale' and `inner scale' of the turbulence
respectively.

Assuming the fluctuations in electron density and magnetic field to be zero
mean isotropic Gaussian random processes with the same outer scale ($l_0$),
\citet{MINTER1996} derived $D_{RM} \propto (\Delta \theta)^{\alpha -2}$.
For three dimensional Kolmogorov turbulence $\alpha=11/3 $, and $D_{RM} \propto
(\Delta \theta)^{5/3}$. However, from Fig.~\ref{structure-function}, we find the
structure function is well fitted by a power law of index 0.7$\pm 0.1$ (Sect.~2.1.1)
up to about 0.7\dg\ and then it gets saturated. This is consistent with
$D_{RM} \propto (\Delta \theta)^{2/3}$, which would indicate $\alpha=8/3 $,
expected from two-dimensional Kolmogorov turbulence. Two-dimensional
turbulence results if the screen responsible for it is confined in thin sheets
in the sky plane.  
\citet{MINTER1996} have found in their data that the structure function slope
changes from about 5/3 to 2/3 at a length scale of about 7 pc. However, in our
data we do not observe any significant deviation from the fit at the smallest
angular separation in Fig.~\ref{structure-function} near 0.002\dg\
corresponding to a linear scale of 0.3 pc at a distance of the GC. This will
indicate if the turbulence is indeed Gaussian in nature, the thickness of the
screen/screens is $\le$0.3 pc in the central kpc of the Galaxy. On the other
hand, turbulent processes could be non Gaussian.  \citet{BOLDYREV2005} have
shown that the Levy distribution of irregularities in a three-dimensional screen
could explain the shapes and the the scaling of observational pulse profile of
a pulsar. A physical realisation of such a Faraday screen is random
discontinuity in the distribution of electron density and magnetic field. This
process has a divergent second moment and could explain results that otherwise
would require the turbulence to be two-dimensional if Gaussian random process
is assumed.

To estimate the strength of the random magnetic fields, we assume the RMs to be
correlated within the outer scale of the RM structure function (40 pc)
(henceforth called cells).
It is quite easy to show that along our LOS
$$ D_{\tmop{RM}} =\{0.8 \times (\Delta n_e \times < B_{\|} > + n_e \times
\Delta B_{\|}) \times l_0 \times \sqrt{n} \}^2$$ where $\Delta n_e$ and
$\Delta B_{\|}$ correspond to the fluctuating component of the electron density 
and magnetic fields along our LOS respectively. In the above
equation, $l_0$ is the size of each cell and `n' the number of such cells
along each LOS.  As discussed in Sect.~3.2.1, $n_e$ is estimated to be
about 0.4 cm$^-3$, and $\Delta n_e$ is also believed to be about the same. In
Sect.~3.2.1,  $<B_{\|} >$ is estimated to be 0.6 $\mu$G. If the central 2 kpc
region is believed to be responsible for the observed RMs, then there will be
about 50 cells along each LOS. The estimated random magnetic fields
at length scales of 40 pc corresponding to the RM structure function of
3.7$\times$10$^5$ \rad\ (Fig.~\ref{structure-function}) is 6~$\mu$G.
We note that electron density distribution is quite
clumpy in the inner Galaxy \citep{CORDES1985}. Therefore, the number of such
cells could be much less, such that the total dispersion measure remains almost
the same. In that case, $\Delta B_{\|}$ would be given by $\sim 6\times
\sqrt{(n/50)}~\mu$G.

\subsection{Implications for the GC magnetic field}

In previous sections we have estimated an average LOS systematic magnetic field
of $\sim$1$~\mu$G and a random field of 6 $\mu$G.  However, this does not
address the overall magnetic field in the central one degree from the GC, which
is described below.

The observed magnetic fields in galaxies are rarely systematic. This is due to
turbulence, and the random component has a field strength that is about the
same in magnitude as the systematic field \citep{ZWEIBEL1997}.  In the GC
region, a highly turbulent magnetoionic media causes high scatter broadening of
extragalactic sources. 
Here we estimate the strength of this random component in this region from our
data, which will provide an estimate of the strength of the systematic field.
As shown in Sect.~2.1.1, the outer scale of RM of sources in this region is
about 40 pc, and we follow the same approach as in the previous section for
calculating the random magnetic fields.
There are about 7 cells within a region of angular radius 1\dg\ corresponding
to a linear size of about 300 pc at a distance of 8.0 kpc.  With an electron
density of 0.4 cm$^{-3}$ in the region, if there is a net LOS magnetic field
of 1 milliGauss over a size scale equivalent to the size of these cells, this
region would introduce a RM of $\sim$12800 \rad.  As the magnetic fields in
these cells are uncorrelated, the mean value of RMs towards sources could be
small, but the rms value of RMs along different LOSs would be $\sim$34,000
\rad. There are 6 source components in our sample seen through the
central 1.1\dg\ from the GC, but we do not find any of their absolute RMs to
be significantly higher than the mean RM from the whole sample. The estimated
rms RM from our sample is consistent with a random field of $\sim$20 $\mu$G in
this central 300 pc region of the Galaxy.  
This suggests that the strong magnetic fields near the NTFs could only be a local
enhancement to the GC magnetic fields and does not fill the entire 300 pc
region.

This outer scale is much larger than the measured size scale of the Faraday
screen of $\sim10^{''}$ (0.4 pc) towards the GC NTFs G359.54+0.18
\citep{YUSEF-ZADEH1997} and Snake (G359.1$-$0.2) \citep{GRAY1995}, and is
consistent with the size scale of low-density HII regions in the Galaxy
\citep{ANANTHARAMAIAH1985}. This is also close to the turbulence scale 
expected from supernova explosions in the GC \citep{BOLDYREV2006}. The 
difference of two order of magnitude in the size scale of turbulence in the
Faraday RMs towards the background sources as compared to regions close to NTFs
indicates the magnetoionic properties of ISM in the GC region is vastly
different than what is observed close to NTFs.  Recent observations
\citep{LAROSA2005} and a model \citep{BOLDYREV2006} also support this
conclusion. If NTFs are dynamic structures, local enhancement of magnetic
fields in their vicinity could also be explained \citep{SHORE1999}.

\section{Conclusions}
\label{rm.summary}
To study the properties of the Faraday screen near the GC, we measured
RMs towards 60 background extragalactic sources through the
$-$6$^{\circ}<l<$6$^{\circ}$, $-2^{\circ}<b<$2$^{\circ}$ region of the Galaxy.
To our knowledge, this provides the first direct determination of large-scale
magnetoionic properties of the central 1 kpc region of the Galaxy not biased by
NTF environments. 
We find a large-scale LOS magnetic fields that point towards us. Either the
bisymmetric spiral model of magnetic field in the Galaxy or the magnetic-field
lines that are mostly aligned with the central bar of the Galaxy could explain
a largely positive RM in the central 1 kpc of the Galaxy.
This large-scale magnetic field has a lower limit of 0.6 $\mu$G along the LOS.
The outer scale of the RM structure function is about 40 pc. 
The RM structure function is well-fitted with a power law index of 0.7$\pm$0.1 at
length scales of 0.3 to 100 pc at the distance of the GC, which is
inconsistent with a three dimensional Kolmogorov turbulence.  A magnetic field
fluctuation of $\sim 6~\mu$G along with electron density fluctuation could
explain the observed RM structure function in the central 1 kpc of the Galaxy.
However, in the inner 300 pc, the maximum random component of the magnetic
field is estimated to be $\sim$20~$\mu$G.  Since GC region has a highly
turbulent ISM, this random magnetic field is very likely have a similar
strength to the systematic field. The observed outer scale of the
magnetoionic medium in this region also does not appear to be less than what is
determined from the whole sample ($\sim$40 pc). This is much larger than the
scale size of the RM structure function $\sim$10$''$ (0.4 pc) observed near the
NTFs in the GC. This indicates that properties of the Faraday screen in the GC is
very different from what is found close to the NTFs. The milliGauss magnetic
fields estimated near the NTFs are localised and do not pervade the central
300 pc of the Galaxy. A more detailed investigation of the magnetic field
involving background sources several times more than the present study would,
however, be required to make a model of the magnetic field configuration in
the region.

\begin{acknowledgements}
	We thank Rajaram Nityananda for introducing the Bootstrap technique to
	us. We also thank the anonymous referee whose comments helped to
	improve the quality of the paper.
\end{acknowledgements}

\bibliographystyle{aa}
\bibliography{gc.rm.pap}
\label{lastpage}
\end{document}